\documentclass[runningheads]{svmult}

\usepackage{makeidx}   % allows index generation
\usepackage{graphicx}  % standard LaTeX graphics tool
\usepackage{subeqnar}  % subnumbers individual equations
\usepackage{multicol}  % used for the two-column index
\usepackage{physprbb}  % modified textarea for proceedings,
\makeindex             % used for the subject index
\begin{document}
\title*{Fact: Many {\rm SCUBA} galaxies harbour AGNs}
\toctitle{Fact: Many {\rm SCUBA} galaxies harbour AGNs}
% allows explicit linebreak for the table of content
%
%
\titlerunning{Fact: Many {\rm SCUBA} galaxies harbour AGNs}
% allows abbreviation of title, if the full title is too long
% to fit in the running head
%

%
\author{D.M.~Alexander\inst{1}
\and F.E.~Bauer\inst{1}
\and S.C.~Chapman\inst{2} 
\and I.~Smail\inst{3} 
\and A.W.~Blain\inst{2}
\and W.N.~Brandt\inst{4}
\and R.J.~Ivison\inst{5}}

\authorrunning{ALEXANDER ET~AL.}

\institute{Institute of Astronomy, Madingley Road, Cambridge CB3 0HA,
  UK 
\and California Institute of Technology, Pasadena, California 91125,
  USA
\and Institute for Computational Cosmology, University of Durham,
  South Road, Durham DH1 3LE, UK
\and Department of Astronomy and Astrophysics, Pennsylvania
  State University, 525 Davey Laboratory, University Park, PA 16802,
  USA
\and Astronomy Technology Centre, Royal Observatory, Blackford
  Hill, Edinburgh EH9 3HJ, UK}

% if there are more than two authors,
% please abbreviate author list for running head
%
%
%\institute{Instiute of Astronomy, Cambridge, CB3 0HA, UK}
%\and Universit\'{e} de Paris-Sud,
%     Laboratoire d'Analyse Num\'{e}rique,
%     B\^{a}timent 425,\\
%     F-91405 Orsay Cedex, France}

\maketitle              % typesets the title of the contribution

\begin{abstract}
  
  Deep {\rm SCUBA} surveys have uncovered a large population of
  ultra-luminous galaxies at $z>1$. These sources are often assumed to
  be starburst galaxies, but there is growing evidence that a
  substantial fraction host an AGN (i.e., an accreting super-massive
  black hole). We present here possibly the strongest evidence for
  this viewpoint to date: the combination of ultra-deep X-ray
  observations (the 2~Ms Chandra Deep Field-North) and deep optical
  spectroscopic data. We argue that upward of 38\% of bright ($f_{\rm
    850\mu m}\ge$~5~mJy) {\rm SCUBA} galaxies host an AGN, a fraction
  of which are obscured QSOs (i.e.,\ $L_X>
  3\times10^{44}$~erg~s$^{-1}$). However, using evidence from a
  variety of analyses, we argue that in almost all cases the AGNs are
  not bolometrically important (i.e.,\ $<20$\%).  Thus, star formation
  appears to dominate their bolometric output. A substantial fraction
  of bright {\rm SCUBA} galaxies show evidence for binary AGN
  activity. Since these systems appear to be interacting and merging
  at optical/near-IR wavelengths, their super-massive black holes will
  eventually coalesce.

\end{abstract}

\section{Introduction}

Blank-field {\rm SCUBA} surveys have uncovered a large population of
submillimetre (submm; $\lambda$~=~300--1000~$\mu$m) emitting galaxies
($\approx$~1000--10000 sources deg$^{-2}$ at $f_{\rm 850\mu
  m}\approx$~\hbox{1--5~mJy}; e.g.,\ [5,12,13,22,27,39]). The
majority of these sources are faint at all other wavelengths,
hindering source identification studies. However, due to a
considerable amount of intensive multi-wavelength follow-up effort, it
is becoming clear that almost all are dust-enshrouded galaxies at
$z>1$ (e.g.,\ [17,30,41,47]). With estimated bolometric luminosities
of $\approx 10^{12}$--$10^{13}$ $\rm L_{\odot}$, these galaxies
outnumber comparably luminous local galaxies by several orders of
magnitude.

Central to the study of submm galaxies is the physical origin of their
extreme luminosities (i.e.,\ starburst or AGN activity). If these
sources are shown to be ultra-luminous starburst galaxies then their
derived star-formation rates suggest a huge increase in star-formation
activity at $z>1$. Conversely, if these sources are shown to be
ultra-luminous AGNs then they will outnumber comparably luminous
optical QSOs by $\approx$~1--2 orders of magnitude.  Both of these
scenarios provide challenges to models of galaxy formation and
evolution.

\section{Searching for AGN activity in {\rm SCUBA} galaxies}

It was initially expected that deep {\rm SCUBA} surveys would identify
large numbers of high-redshift ultra-luminous starburst galaxies.
However, the first {\rm SCUBA} source to be unambiguously identified
via optical/near-IR spectroscopy was actually found to contain an
obscured AGN, probably a BALQSO (i.e.,\ the $z=2.81$ source
SMMJ02399--0136; [28,44]); subsequent moderately deep {\it Chandra}
observations showed that this AGN is luminous at \hbox{X-ray} energies
($L_X\approx 10^{45}$~erg~s$^{-1}$; [11]). Although clearly containing
a powerful AGN, sensitive CO (3--2) observations of SMMJ02399--0136
suggested that star formation contributes $\approx$~50\% of the total
bolometric luminosity [25]. The lesson learnt from SMMJ02399--0136 is
that the identification of an AGN at optical/near-IR wavelengths does
not necessarily imply that the AGN is bolometrically dominant.

Extensive optical follow-up observations of other {\rm SCUBA} sources
have revealed further AGNs, a few starburst galaxies, and many sources
with uncertain classifications (e.g.,\ [17,29,41]). Although clearly
successful in identifying the presence of an AGN in some {\rm SCUBA}
galaxies, optical spectroscopy has a number of limitations for AGN
identification in all {\rm SCUBA} galaxies:

\begin{itemize}
  
\item The large positional uncertainty of {\rm SCUBA} sources often
  makes it challenging to identify an optical counterpart securely.
  
\item A large fraction of the {\rm SCUBA} sources with secure optical
  counterparts are optically faint ($I\ge24$), making optical source
  classification challenging despite success in redshift
  identification (e.g.,\ [17]).
  
\item AGNs do not always show clear AGN signatures at optical
  wavelengths due to dust absorption and dilution by host-galaxy light
  (e.g.,\ [20]).

\end{itemize}

Arguably the best discriminator of AGN activity is the detection of
luminous hard X-ray emission (i.e.,\ $>$~2~keV).\footnote{Radio
  observations can also be useful in identifying AGNs in {\rm SCUBA}
  galaxies; however, not all AGNs are bright at radio wavelengths, and
  radio emission from strong star formation can easily mask the radio
  emission from a radio-quiet AGN.} Hard X-ray emission is relatively
insensitive to obscuration (at least for sources that are Compton
thin; i.e.,\ $N_{\rm H}< 1.5\times 10^{24}$~cm$^{-2}$), and any hard
X-ray emission from star formation in the host galaxy is often
insignificant when compared to that produced by the AGN. Hard X-ray
observations can even provide a secure AGN identification in sources
where the optical signatures and counterparts are weak or even non
existent (e.g.,\ [1]). Due to their high-energy X-ray coverage, high
X-ray sensitivity, and excellent positional accuracy, the {\it
  Chandra} and {\it XMM-Newton} observatories offer the best
opportunities for the X-ray investigation of {\rm SCUBA} galaxies.

\section{The promise of {\it Chandra} and {\it XMM-Newton}}

The first cross-correlation studies of moderately deep X-ray surveys
with deep {\rm SCUBA} surveys yielded no overlap between the X-ray and
submm detected source populations ($<$10--20\%;
[24,26,38]).\footnote{Further cross-correlation studies with
  moderately deep X-ray observations have revealed some overlap
  [4,6,30,46]. This is probably due to the larger areal coverage of
  the {\rm SCUBA} observations.} This was somewhat contrary to
expectations given that some submm galaxies clearly host an AGN.
However, in retrospect, considering that the two most obvious AGNs
identified via optical spectroscopy were both detected by {\it
  Chandra} (i.e.,\ SMMJ02399--0134 and SMMJ02399--0136; [11,41]), AGN
identification with moderately-deep X-ray observations had not
performed any worse than that achieved via optical spectroscopy.
These early studies concluded that a reasonable fraction of
bolometrically dominant AGNs can only be present in the submm galaxy
population if they are obscured by Compton-thick material. In this
scenario, almost all of the direct emission from the AGN is obscured
at X-ray wavelengths.

Later studies with the Chandra Deep Field-North (CDF-N; 1~Ms exposure:
[15], 2~Ms exposure: [2]) survey showed that a significant fraction of
submm galaxies are detected in ultra-deep X-ray observations [3,7].
In the 2~Ms {\it Chandra} exposure, seven ($>36$\%) bright {\rm SCUBA}
galaxies ($f_{\rm 850\mu m}\ge$~5~mJy; S/N$>$4) had an X-ray
counterpart in a 70.3 arcmin$^2$ region centred around the CDF-N
aim-point [3].\footnote{The X-ray sources were matched to
  radio-detected {\rm SCUBA} and radio-undetected {\rm SCUBA} sources
  using 1$^{\prime\prime}$ and 4$^{\prime\prime}$ search radii,
  respectively; the respective probabilities of projected chance
  associations are $<$1\% and 4\% per {\rm SCUBA} source.} At the time
of this study, complete {\rm SCUBA} observations across the 70.3
arcmin$^2$ region were not available and only a lower limit on the
X-ray-submm fraction could be placed. The X-ray detected submm galaxy
fraction is 54\% when the most complete {\rm SCUBA} observations of
[14] are used.

Figure~1 shows the X-ray detected submm galaxy source density versus
\hbox{X-ray} flux for the X-ray-submm cross-correlation studies to
date.  Clearly, the detection of significant numbers of submm galaxies
at X-ray energies requires deep or wide-area observations. For
example, $\approx$~15 ($\approx$~10--20\%) {\rm SCUBA} sources should
have X-ray counterparts in the forthcoming 0.25 deg$^2$ SHADES survey
(P.I. J. Dunlop) of the {\it XMM-Newton} Subaru Deep Survey (XMM-SDS)
region, even though the X-ray coverage is only moderately deep
(50--100~ks {\it XMM-Newton} exposures; P.I. M.  Watson).

%
% X-ray-submm source densities
%
\begin{figure}[t]
\begin{center}
\includegraphics[width=.6\textwidth]{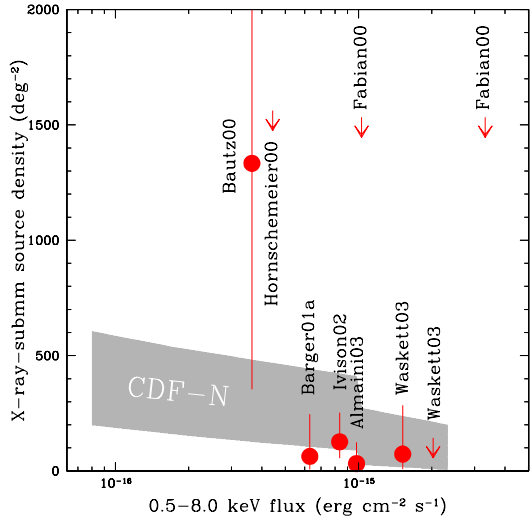}
\end{center}
\caption[]{X-ray detected submm galaxy source density versus X-ray flux. The shaded region corresponds to the constraints from the CDF-N studies. The data points and error bars correspond to the constraints placed by different X-ray-submm cross-correlation studies (see references); the arrows indicate upper limits. The flux limits are calculated assuming $\Gamma=1.0$, and a 5 ({\it Chandra}) or 20 ({\it XMM-Newton}) count detection threshold for each X-ray survey; the flux limits are corrected for gravitational lensing amplification where necessary.}
\label{eps1}
\end{figure}

\section{The fraction of bright {\rm SCUBA} galaxies hosting an AGN}

On the basis of radio constraints, the expected X-ray luminosity from
star formation in bright submm galaxies is
$L_X=10^{42}$--$10^{43}$~erg~s$^{-1}$ [3,9]. Since it is possible to
detect sources of this luminosity out to $z\approx$~1--3 with the 2~Ms
CDF-N survey, the detection of X-ray emission from a submm galaxy does
not necessarily imply that it hosts an AGN. Indeed, two of the seven
X-ray detected submm galaxies in [3] have X-ray properties consistent
with those expected from star formation activity (i.e.,\ soft and
comparatively low luminosity X-ray emission that is correlated with
the radio emission; see \S3.2 in [3]). With X-ray luminosities of
\hbox{$\approx 4\times10^{42}$~erg~s$^{-1}$}, these would be the most
X-ray luminous starburst galaxies known. The other five X-ray detected
submm galaxies have X-ray properties consistent with those of obscured
AGNs [i.e.,\ hard ($\Gamma<1.0$), luminous
(\hbox{$L_X>10^{43}$~erg~s$^{-1}$}) X-ray emission that is in excess
of that expected from star formation]. Only one of these sources would
have been classified as an AGN based on its optical properties,
underlining the potency of ultra-deep X-ray observations in AGN
identification. On the basis of these classifications at least 38\% of
bright ($f_{\rm 850\mu m}\ge$~5~mJy) {\rm SCUBA} galaxies host AGNs.

Interestingly, all of the radio-detected submm galaxies are X-ray
detected while relatively few of the radio-undetected submm galaxies
are X-ray detected. This X-ray-radio dichotomy is unlikely to be due
to AGN activity since in almost all cases the radio emission appears
to be dominated by star formation. Possible explanations for the
X-ray-radio dichotomy are

\begin{enumerate}
  
\item Some of the radio-undetected submm galaxies may be spurious.
  Due to the steep number counts at submm wavelengths most {\rm SCUBA}
  sources are detected at a comparatively low significance.
  
\item The radio emission from the radio-undetected submm galaxies may
  be extended and therefore ``resolved out'' in the radio
  observations.
    
\item The radio-undetected submm galaxies might contain a cooler dust
  component, detected at submm wavelengths but corresponding to lower
  bolometric and radio luminosities when compared to hotter
  star-formation regions (e.g.,\ cirrus emission; [13,23]).
  
\item The radio-undetected submm galaxies may be essentially the same
  as the radio-detected submm galaxies but lie at higher redshifts.

\end{enumerate}
  
A combination of these effects might explain the X-ray-radio dichotomy
of submm galaxies. Explanation 1 is unlikely to affect significantly
the results of [3] since only {\rm SCUBA} sources with S/N$>$4 were
investigated. Explanations 2 and 3 would imply that the
radio-undetected submm galaxies are different beasts to the
radio-detected submm galaxies and might not typically host AGNs.
Explanation 4 justifies the non detection of the radio-undetected
submm galaxies at X-ray and radio wavelengths as due to a lack of
sensitivity. Under this assumption in particular, the fraction of
bright {\rm SCUBA} galaxies hosting AGNs could be considerably higher
than 38\% (i.e.,\ $\approx$~75\% if explanation 4 is correct and half
of all bright submm galaxies are radio detected).

The closest local analogs to {\rm SCUBA} galaxies are ultra-luminous
infrared galaxies (ULIRGs: $L_{\rm 8-1000\mu m}>$~10$^{12}$ $\rm
L_{\odot}$; [37]). Based on spectroscopic classifications (optical:
[43], mid-IR: [32]), it has been found that the fraction of ULIRGs
hosting an AGN increases with luminosity [from $\approx$~25\% ($L_{\rm
  8-1000\mu m}=$~\hbox{10$^{12}$--10$^{12.3}$} $\rm L_{\odot}$) to
$\approx$~50\% ($L_{\rm 8-1000\mu m}>$~10$^{12.3}$ $\rm L_{\odot}$)].
These AGN fractions are generally consistent with that found here for
bright {\rm SCUBA} galaxies.

%
% Simple band-ratio diagnostics
%
\begin{figure}[t]
\begin{center}
\includegraphics[width=.6\textwidth]{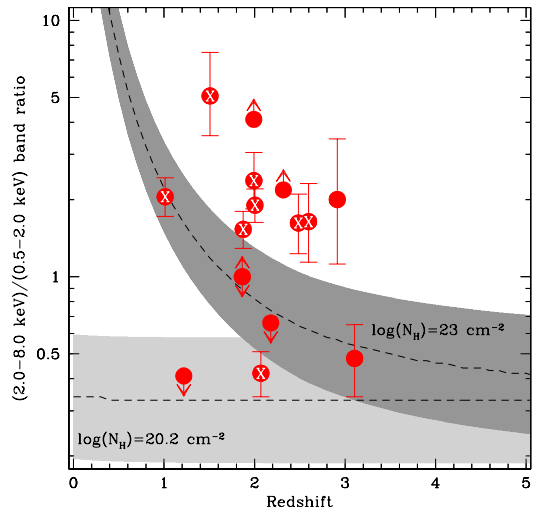}
\end{center}
\caption[]{X-ray band ratio versus spectroscopic redshift for the X-ray detected submm galaxies. The light and dark shaded regions show the range of expected band ratios for an unabsorbed and absorbed AGN, respectively. These regions were calculated assuming a $\Gamma=1.7\pm0.5$ power law with differing amounts of absorption (as shown). This simple figure has limited diagnostic utility; however, it suggests that almost all of the sources are absorbed. The sources indicated by an ``X'' are further investigated using X-ray spectral analyses; see Figure~3.}
\label{eps1}
\end{figure}

%
% Rest-frame spectral slope diagnostics
%
\begin{figure}[t]
\begin{center}
  \includegraphics[width=.6\textwidth]{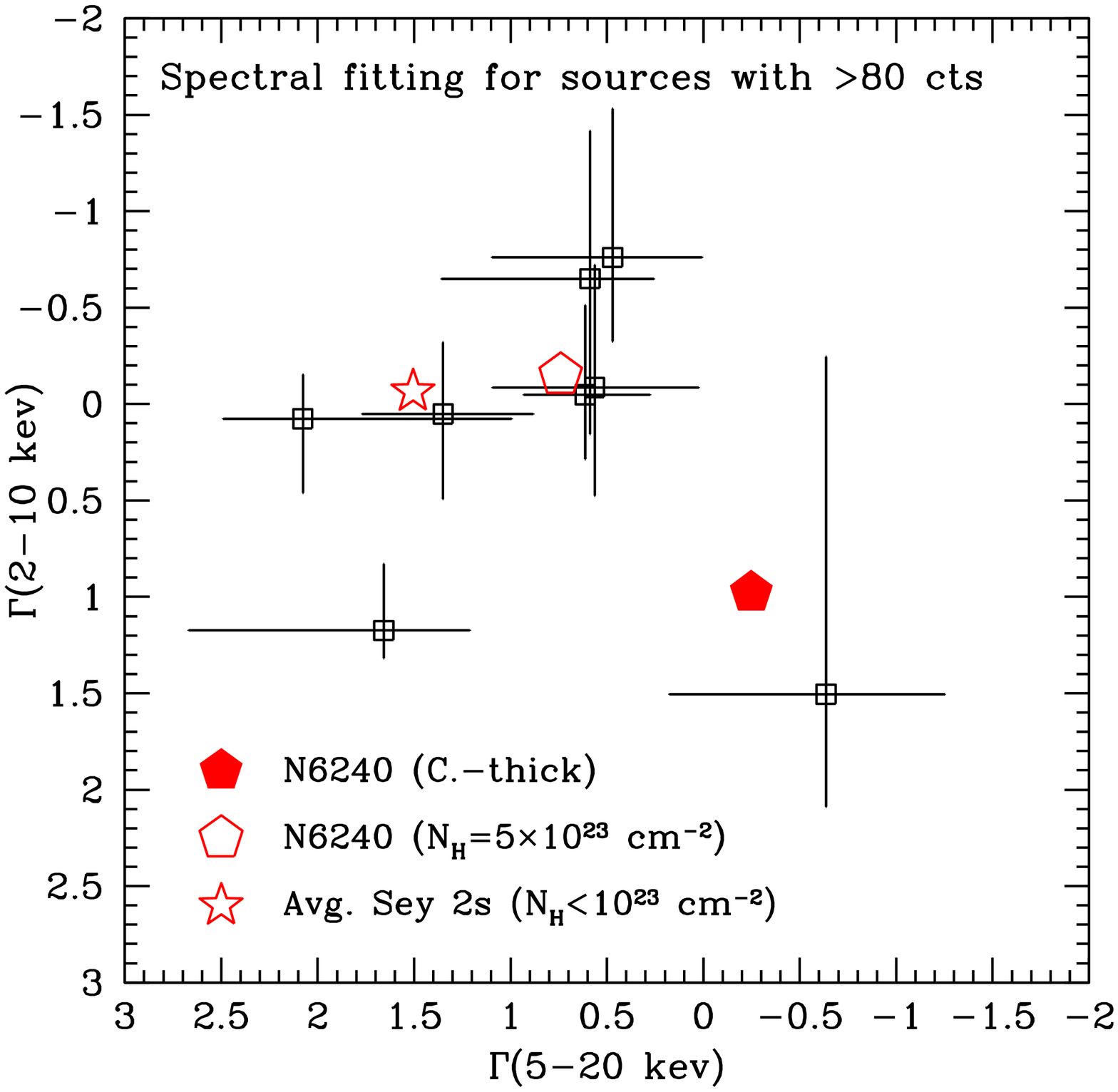}
\end{center}
\caption[]{Rest-frame 2--10~keV versus 5--20~keV spectral slopes for the X-ray brightest ($>$80 counts) submm galaxies compared to nearby absorbed AGNs. The X-ray spectral properties for the nearby absorbed AGNs consist of the average for a sample of Seyfert 2 galaxies (with $N_{\rm H}< 10^{23}$~cm$^{-2}$, open star; [33]), NGC~6240 adjusted as though it was Compton thin (with $N_{\rm H}= 5\times10^{23}$~cm$^{-2}$, open pentagon) and NGC~6240 (i.e.,\ Compton thick, filled pentagon; [45]). The X-ray-submm galaxies generally reside in the region found for Compton-thin AGNs.}
\label{eps1}
\end{figure}

\section{What powers bright {\rm SCUBA} galaxies?}

The ultra-deep CDF-N studies showed that a significant fraction of
submm galaxies host obscured AGNs. As discussed in \S2 and \S3, in
order to determine if these AGNs are bolometrically dominant it is
crucial to determine if the obscuration toward the AGN is Compton
thick or Compton thin. The most direct discrimination between
Compton-thick and Compton-thin absorption is made with X-ray spectral
analyses. The X-ray spectrum of a Compton-thick AGN is generally
characterised by a large equivalent width Iron K$\alpha$ emission line
($EW\ge$~1~keV; e.g.,\ [8,34]) and a flat or inverted ($\Gamma<0$)
X-ray spectral slope, due to pure reflection.\footnote{A minority of
  Compton-thick AGNs have $EW\approx$~0.5--1.0~keV [8].} By contrast,
the X-ray spectrum of a Compton-thin AGN is usually well fitted by an
absorbed power-law model and a smaller equivalent width Iron K$\alpha$
emission line (generally $EW\approx$~0.1--0.5~keV; e.g.,\ [8,35]).

Basic X-ray spectral analyses were performed on the five submm
galaxies hosting AGNs in the 2~Ms CDF-N study of [3]. Three of the
sources showed the characteristics of Compton-thin absorption, one
source was likely to be Compton thick, and the constraints for the
other source were poor. A comparison of the X-ray-to-submm spectral
slopes of these submm galaxies to those of three nearby luminous
galaxies (Arp~220, a starburst galaxy; NGC~6240, an obscured AGN;
3C273, a quasar) suggested that the AGNs contributed only a small
fraction of the bolometric luminosity (i.e.,\ a few percent). However,
although the study of [3] provided the tightest X-ray constraints on
submm galaxies to date, only one source had a spectroscopic redshift
[the rest of the sources had redshifts determined with the
considerably less certain radio-submm photometric redshift technique
(e.g.,\ [16]), restricting more accurate and quantitative conclusions].

\subsection{The X-ray properties of AGNs in bright {\rm SCUBA} galaxies}

Considerable progress in the optical identification of {\rm SCUBA}
galaxies has been recently made due to the pioneering deep optical
spectroscopic work of [17,19]. By targeting radio and/or X-ray
detected {\rm SCUBA} galaxies, source redshifts for a sizable fraction
of the submm galaxy population have now been obtained. CO emission
line observations have confirmed that both the redshift and
counterpart are correct in many cases (e.g.,\ [36]; R.~Genzel et~al.,
these proceedings). The CDF-N field was one of the fields targeted for
this intensive spectroscopic follow-up: optical spectroscopic
redshifts have been obtained for 24 {\rm SCUBA} galaxies in the CDF-N.
The combination of this deep spectroscopic data with the 2~Ms CDF-N
observations provides powerful constraints on AGNs in submm galaxies.
In particular, reliable spectroscopic redshifts improve the accuracy
of the X-ray spectral analyses through the identification of subtle
X-ray spectral features and making comparisons in rest-frame energy
bands. In these analyses we will focus on the $z>1$ sources, which are
more typical of the general submm galaxy population [17].

Fifteen of the 20 $z>1$ submm galaxies have X-ray counterparts: 12
($\approx$~75\%) of the 16 $z>1$ radio-detected submm galaxies have
X-ray counterparts, continuing the X-ray-radio trend (see \S4). Since
some of the X-ray detected submm galaxies do not have enough counts
for X-ray spectral analyses, we will first compare their X-ray band
ratios [defined as the ratio of the hard-band (2--8~keV) to soft-band
(0.5--2~keV) count rate] to those expected from a simple AGN model;
see Figure~2. This figure has limited diagnostic utility because AGNs
often have more complex spectra than that of power-law emission with
differing amounts of absorption; however, it suggests that few of the
sources are unabsorbed.

We have performed more detailed X-ray spectral analyses for the eight
submm galaxies with $>80$ X-ray counts; the X-ray spectra were
extracted following the procedure outlined in [10]. In Figure~3 we
compare their fitted rest-frame spectral slopes (in the 2--10~keV and
5--20~keV bands) to those found for nearby AGNs with differing amounts
of absorption. Although this is still a relatively crude diagnostic,
there are distinctions between Compton-thin and Compton-thick sources.
Based on this analysis, the majority of the sources appear to be
heavily obscured but still only Compton thin (i.e.,\ $N_{\rm
  H}\approx$~few~$\times10^{23}$~cm$^{-2}$). We also searched for the
presence of Iron K$\alpha$ emission lines. The rest-frame
equivalent-width constraints are generally quite weak: Iron K$\alpha$
emission lines are possibly detected in two sources (with
$EW\approx$~0.7~keV and $EW\approx$~1.7~keV) while the other six only
have upper limits (all have $EW<$~1.8~keV and five have
$EW<$~1.0~keV). From these analyses it appears unlikely that more than
$\simeq$~3 of these 8 submm galaxies contain a Compton-thick AGN. We
cannot say much about the individual X-ray properties of the seven
X-ray-submm galaxies with $<80$ X-ray counts; however, we note that
since their band ratios are consistent with the sources for which we
have performed X-ray spectral analyses, they probably have similar
amounts of absorption (see Figure~2).

%
% Far-IR versus X-ray luminosity comparison
%
\begin{figure}[t]
\begin{center}
\includegraphics[width=.6\textwidth]{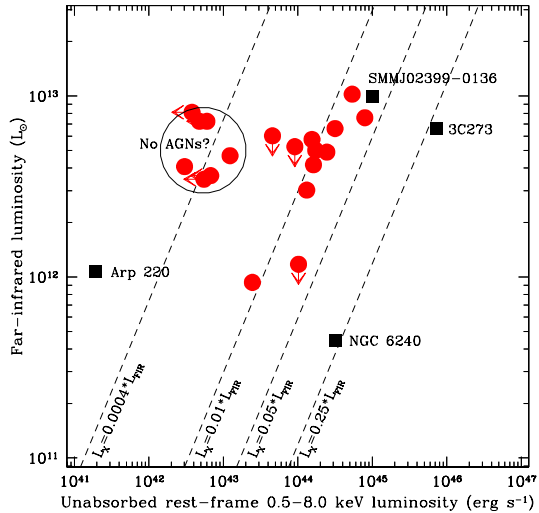}
\end{center}
\caption[]{Rest-frame far-IR versus unabsorbed 0.5--8.0~keV X-ray luminosities for the X-ray detected submm galaxies with spectroscopic redshifts. The far-IR luminosities have been calculated from the 1.4~GHz radio luminosity (assuming the far-IR-radio correlation) and the X-ray luminosities have been corrected for the effect of absorption. The submm galaxy SMMJ02399--0136, and three luminous nearby galaxies (Arp~220, NGC~6240, and 3C273) are shown for comparison. The line indicating the smallest $L_X/L_{FIR}$ ratio corresponds to the average found for starburst galaxies (e.g.,\ [9]).}
\label{eps1}
\end{figure}

\subsection{The bolometric AGN contribution to bright {\rm SCUBA} galaxies}

Based on these analyses, the full range of unabsorbed X-ray
luminosities is \hbox{$L_X\approx 10^{42}$--$10^{45}$~erg~s$^{-1}$};
see Figure~4.\footnote{The corrections for absorption are generally a
  factor of $\approx$~3.} While a number of the sources have X-ray
luminosities consistent with those of X-ray luminous starburst
galaxies (i.e.,\ no AGNs), the majority clearly host AGNs. The AGNs
generally have X-ray luminosities consistent with those of Seyfert
galaxies; however, three could be considered obscured QSOs (i.e.,\ 
$L_X> 3\times10^{44}$~erg~s$^{-1}$).

We calculated rest-frame far-IR luminosities for all of the sources,
following Equation 2 in [3] and assuming the local radio-far-IR
correlation. The comparison of rest-frame far-IR luminosity with
unabsorbed X-ray luminosity is shown in Figure~4. This figure provides
an indicator of the AGN contribution to the bolometric
luminosity.\footnote{When the radio emission has a large AGN component
  the far-IR luminosity will be overestimated and the AGN bolometric
  contribution will be underestimated; however, in general the radio
  emission appears to be star-formation dominated.}  Assuming that the
far-IR emission from NGC~6240 and 3C273 is dominated by AGN activity,
the AGNs in these submm galaxies contribute at most 20\% of the
bolometric luminosity and more typically a few percent. If instead we
determine the AGN bolometric contributions based on the spectral
energy distribution of SMMJ02399--0136 (i.e.,\ $\approx$~50\%:
[11,25]; see \S2) then the AGN contributions increase by a factor of
$\approx$~2.5. Clearly, there is a range of X-ray to bolometric
luminosity conversions for AGNs; however, on average the AGNs are
unlikely to contribute more than $\approx$~10--20\% of the bolometric
luminosity. Hence, although a large fraction of bright {\rm SCUBA}
galaxies host an AGN (i.e.,\ at least $\approx$~38\%), in general,
star formation is likely to dominate their bolometric output.

\section{Evidence for binary AGN activity}

An unexpected result in the 2~Ms study of [3] was that two
($\approx$~30\%) of the seven X-ray detected submm galaxies were
individually associated with X-ray pairs. The small angular
separations of these pairs ($\approx$~2--3$^{\prime\prime}$)
correspond to just $\approx$~20~kpc at $z=2$ (approximately one
galactic diameter); the probability of a projected chance association
is $<$1\%. From {\it HST} imaging, it has been shown that the majority
of {\rm SCUBA} sources appear to be galaxies involved in major mergers
(e.g.,\ [18,21,29,40]). At X-ray energies we are presumably witnessing
binary AGN activity fuelled by galaxy mergers that will ultimately
lead to the coalescence of the super-massive black holes. Recent
X-ray studies are only just beginning to show the late stages of
binary AGN activity in AGNs at much lower redshifts (e.g.,\ NGC~6240;
[31]); however, in this ultra-deep X-ray observation we have good
evidence that this is occurring at $z\approx2$! Since the smallest
linear separation we can resolve with {\it Chandra} at $z\approx$~2 is
$\approx$~10~kpc, many of the other X-ray detected submm galaxies
could be binary AGNs with smaller separations (e.g.,\ the linear
separation of the two AGNs in NGC~6240 is $\approx$~1~kpc). Five
($\approx$~3\%) of the 193 X-ray sources in this region are close
X-ray pairs, showing that binary AGN behaviour appears to be closely
associated with submm galaxies (see also [42]). Qualitatively, this
picture is consistent with that expected for major merger activity.

%%%%%%%%%%%%%%%%%%%%%%%%%%%%%%%%%%%%%%%%%%%%%%%%%%%%%%%%%%%%%%%%%%%%%%

%INDEX%%%%%%%%%%%%%%%%%%%%%%%%%%%%%%%%%%%%%%%%%%%%%%%%%%%%%%%%%%%%%%%
% Please check with the editor of your book whether he plans to
% include a "mutual" subject index - if so, please code your entries
% in the standard syntax. For your own purposes you may print your
% "personal" index by using the following commands:
%
%\clearpage
%\addcontentsline{toc}{section}{Index}
%\flushbottom
%\printindex
%%%%%%%%%%%%%%%%%%%%%%%%%%%%%%%%%%%%%%%%%%%%%%%%%%%%%%%%%%%%%%%%%%%%%


\begin{thebibliography}{8.}
\addcontentsline{toc}{section}{References}

\bibitem{} Alexander, D.~M., Brandt, W.~N., Hornschemeier, A.~E.,
  et~al.\ 2001, AJ, 122, 2156

\bibitem{} Alexander, D.~M., Bauer, F.~E., Brandt, W.~N., et~al.\ 
  2003a, AJ, 126, 539
  
\bibitem{} Alexander, D.~M., Bauer, F.~E., Brandt, W.~N., et~al.\ 
  2003b, AJ, 125, 383
  
\bibitem{} Almaini, O., Scott, S.~E., Dunlop, J.~S., et~al.\ 2003,
  MNRAS, 338 303
  
\bibitem{} Barger, A.~J., Cowie, L.~L., Sanders, D.~B.\ 1999, ApJ,
  518, L5
  
\bibitem{} Barger, A.~J., Cowie, L.~L., Mushotzky, R.~F., et~al.\ 
  2001a, AJ, 121, 662
  
\bibitem{} Barger, A.~J., Cowie, L.~L., Steffen, A.~T., et~al.\ 2001b,
  ApJ, 560, L23
  
\bibitem{} Bassani, L., Dadina, M., Maiolino, R., et~al.\ 1999, ApJS,
  121, 473
  
\bibitem{} Bauer, F.~E., Alexander, D.~M., Brandt, W.~N., et~al.\ 
  2002, AJ, 124, 2351

\bibitem{} Bauer, F.~E., Vignali, C., Alexander, D.~M., et~al.\ 2003,
  Adv. Space Res., in press (astro-ph/0209415)
  
\bibitem{} Bautz, M.~W., Malm, M.~R., Baganoff, F.~K., et~al.\ 2000,
  ApJ, 543, L119
  
\bibitem{} Blain, A.~W., Smail, I., Ivison, R.~J., et~al.\ 1999,
  MNRAS, 302, 632
  
\bibitem{} Blain, A.~W., Smail, I., Ivison, R.~J., et~al.\ 2002,
  Physics Reports, 369, 111
  
\bibitem{} Borys, C., Chapman, S., Halpern, M., et~al.\ 2003, MNRAS,
  344, 385
  
\bibitem{} Brandt, W.~N., Alexander, D.~M., Hornschemeier, A.~E.,
  et~al.\ 2001, AJ, 122, 2810

\bibitem{} Carilli, C.~L., Yun, M.~S.\ 1999, ApJ, 513, L13
  
\bibitem{} Chapman, S.~C., Blain, A.~W., Ivison, R.~J., Smail, I.~R.\ 
  2003a, Nature, 422, 695
  
\bibitem{} Chapman, S.~C., Windhorst, R., Odewahn, S., et~al.\ 2003b,
  ApJ, 599, 92
  
\bibitem{} Chapman, S.~C., et~al.\ 2004, ApJ, submitted
  
\bibitem{} Comastri, A., Mignoli, M., Ciliegi, P., et~al.\ 2002, ApJ,
  571, 771
  
\bibitem{} Conselice, C.~J., Chapman, S.~C., Windhorst, R.~A.\ 2003,
  ApJ, 596, L5
  
\bibitem{} Cowie, L.~L., Barger, A.~J., Kneib, J.-P.\ 2002, AJ, 123,
  2197

\bibitem{} Efstathiou, A., Rowan-Robinson, M.\ 2003, MNRAS, 343, 322 
  
\bibitem{} Fabian, A.~C., Smail, I., Iwasawa, K., et~al.\ 2000, MNRAS,
  315, L8
  
\bibitem{} Frayer, D.~T., Ivison, R.~J., Scoville, N.~Z., et~al.\ 
  1998, ApJ, 506, L7
  
\bibitem{} Hornschemeier, A.~E., Brandt, W.~N., Garmire, G.~P.,
  et~al.\ 2000, ApJ, 541, 49
  
\bibitem{} Hughes, D.~H., Serjeant, S., Dunlop, J., et~al. 1998,
  Nature, 394, 241
 
\bibitem{} Ivison, R.~J., Smail, I., Le Borgne, J.-F., et~al.\ 1998,
  MNRAS, 298, 583
  
\bibitem{} Ivison, R.~J., Smail, I., Barger, A.~J., et~al.\ 2000,
  MNRAS, 315, 209
  
\bibitem{} Ivison, R.~J., Greve, T.~R., Smail, I., et~al.\ 2002,
  MNRAS, 337, 1
  
\bibitem{} Komossa, S., Burwitz, V., Hasinger, G., et~al.\ 2003, ApJ,
  582, L15
  
\bibitem{} Lutz, D., Spoon, H.~W.~W., Rigopoulou, D., et~al.\ 1998,
  ApJ, 505, L103

\bibitem{} Malizia, A., Bassani, L., Stephen, J.~B., et~al.\ 2003,
  ApJ, 589, L17

\bibitem{} Matt, G., Brandt, W.~N., Fabian, A.~C.\ 1996, MNRAS, 280,
  823
  
\bibitem{} Nandra, K., George, I.~M., Mushotzky, R.~F., et~al.\ 1997,
  ApJ, 477, 602
    
\bibitem{} Neri, R.~et al.\ 2003, ApJ, 597, L113 

\bibitem{} Sanders, D.~B.~\& Mirabel, I.~F.\ 1996, ARA\&A, 34, 749

\bibitem{} Severgnini, P., Maiolino, R., Salvati, M., et~al.\ 2000,
  A\&A, 360, 457

\bibitem{} Smail, I., Ivison, R.~J., Blain, A.~W.\ 1997, ApJ, 490, L5
  
\bibitem{} Smail, I., Ivison, R.~J., Blain, A.~W., Kneib, J.-P.\ 
  1998, ApJ, 507, L21

\bibitem{} Smail, I., Ivison, R.~J., Blain, A.~W., et~al.\ 2002,
  MNRAS, 331, 495
  
\bibitem{} Smail, I., Scharf, C.~A., Ivison, R.~J., et~al.\ 2003, ApJ,
  599, 86
  
\bibitem{} Veilleux, S., Kim, D.-C., Sanders, D.~B.\ 1999, ApJ,
  522, 113

\bibitem{} Vernet, J., Cimatti, A.\ 2001, A\&A, 380, 409
  
\bibitem{} Vignati, P., Molendi, S., Matt, G., et~al.\ 1999, A\&A,
  349, L57

\bibitem{} Waskett, T.~J., Eales, S.~A., Gear, W.~K., et~al.\ 2003,
  MNRAS, 341, 1217
  
\bibitem{} Webb, T.~M.~A., Lilly, S.~J., Clements, D.~L., et~al.\ 
  2003, ApJ, 597, 680

\end{thebibliography}
\end{document}